%% file: main.tex
\documentclass[conference]{IEEEtran}
\IEEEoverridecommandlockouts
\usepackage{cite}
\usepackage{amsmath,amssymb,amsfonts,physics,caption,subcaption}
\usepackage{algorithmic}
\usepackage{graphicx}
\usepackage{textcomp}
\usepackage{xcolor}
\usepackage{amsthm}
\newtheorem{theorem}{Theorem}
\newtheorem{corollary}[theorem]{Corollary}
\newtheorem{lemma}[theorem]{Lemma}
\newtheorem{observation}[theorem]{Observation}

\newtheorem{definition}[theorem]{Definition}

\def\BibTeX{{\rm B\kern-.05em{\sc i\kern-.025em b}\kern-.08em
    T\kern-.1667em\lower.7ex\hbox{E}\kern-.125emX}}
\begin{document}
\def\qaoa{\textsc{QAOA}\textsuperscript{+}}
\def\ex#1{\mathbb{E}[#1]}
\def\treg{$2$-regular}
\def\Cn{$\textsc{C}_n$}
\title{Applying the Quantum Alternating Operator Ansatz to the Graph Matching Problem
}

\author{
\IEEEauthorblockN{Sagnik Chatterjee}
\IEEEauthorblockA{\textit{Computer Science Department} \\
\textit{Indraprastha Institute of Information Technology}\\
New Delhi, India \\
sagnikc@iiitd.ac.in}
\and
\IEEEauthorblockN{Debajyoti Bera}
\IEEEauthorblockA{\textit{Computer Science Department} \\
\textit{Indraprastha Institute of Information Technology}\\
New Delhi, India \\
dbera@iiitd.ac.in}
}

\maketitle

\begin{abstract}
The Quantum Alternating Operator Ansatz (QAOA+) framework has recently gained attention due to its ability to solve discrete optimization problems on noisy intermediate-scale quantum (NISQ) devices in a manner that is amenable to derivation of  worst-case guarantees. We design a technique in this framework to tackle a few problems over maximal matchings in graphs. Even though maximum matching is polynomial-time solvable, most counting and sampling versions are \#P-hard.

We design a few algorithms that generates  superpositions over matchings allowing us to sample from them.  In particular, we get a superposition over all possible matchings when given the empty state as input and a superposition over all maximal matchings when given the $W$-states as input.

Our main result is that the expected size of the matchings corresponding to the output states of our QAOA+ algorithm when ran on a 2-regular graph is greater than the expected matching size obtained from a uniform distribution over all matchings. This algorithm uses a $W$-state as input and we prove that this input state is better compared to using the empty matching as the input state.

\end{abstract}


\section{Introduction}
\noindent Quantum Approximate Optimization Algorithms (abbreviated as QAOA) is a class of gate-model algorithms that can be implemented on near-term quantum computers (\cite{farhi2014quantum}). Initially, QAOA was designed to be applied in the context of unconstrained optimization problems (\cite{farhi2014quantum},\cite{farhi2014lin}) but any instance in which QAOA performs better than its classical counterparts is yet to be seen. However, Farhi and Harrow showed that efficiently simulating QAOA for even the lowest depth circuits would collapse the Polynomial Hierarchy (\cite{farhi2016quantum}). This put QAOA as a strong contender at the forefront of the Quantum supremacy debate (\cite{Harrow_2017}) which sparked a renewed interest in the field.\par
\noindent  A major modification to the QAOA framework was given in \cite{Hadfield_2019}, where the framework of QAOA was modified to work with constrained optimization problems by producing only feasible states (with respect to the constraints of the problems) on measurement in the computational basis. The authors termed this new framework as the Quantum Alternating Operator Ansatz (which we abbreviate as \qaoa{}). \par
\noindent We turned our attention on applying the \qaoa{} setup to the matching problem. A matching is a set of edges which are vertex disjoint. Finding a maximum matching in a graph is already known to be solvable in polynomial time \cite{edmonds1965paths} classically. There also exists quantum analogues to the classical algorithms that employ Grover amplifications \cite{ambainis2005quantum}. However, counting problems with respect to matchings are \textsc{\#P}-hard \cite{valiant1979complexity}. Hence, there does not exist efficient deterministic classical algorithms to create a superposition over all distinct matchings with non-zero amplitudes or all maximal matchings with non-zero amplitudes in polynomial time.\par
\noindent In this work we design and apply a \qaoa{} style algorithm to two different input states - a quantum state corresponding to the empty matching, and a quantum state corresponding to a superposition over all matchings of size $1$ with a non-zero amplitude. We obtained the following  results:
\begin{itemize}
    \item Even for $p=1$ and starting from the empty matching, our \qaoa{} algorithm creates a superposition over all distinct matchings with non-zero amplitudes.
    \item Using our \qaoa{} setup and the $\ket{W_1}$ state as the initial state, we can converge to a superposition over maximal matchings in iterations at most twice the input size on expectation.
    \item For \treg{} graphs, we show that the output state of our \qaoa{} setup gives us a better expected matching size compared to the expected matching size from a uniform distribution over all matchings.
    \item For \treg{} graphs, we compare the two initial states and show that using a superposition over all distinct matchings having size $1$, we can obtain a better expected matching size compared to using the empty matching as the initial state.
\end{itemize}
\section{Background: Quantum Alternating Operator Ansatz}
\noindent\qaoa{} style algorithms are applied to combinatorial optimization problems. A combinatorial optimization problem may be formulated in terms of $m$ clauses and an $n$-bit string $z$ (which represents $n$ variables).
\begin{equation}
    C(z) = \sum_{i=1}^{m} C_{i}(z)
\end{equation}
where $C_{i}(z)=1$ if $z$ satisfies $C_{i}(z)$, and $0$ otherwise. Any input string $z$ forms the computational basis $\ket{z}$. Optimizing (maximizing in this case) $C(z)$ refers to finding the $z$ for which the maximum number of clauses is satisfied. The objective function is $C(z)$($C:D\rightarrow \mathbb{R}$) which we have to optimize. We consider a Hilbert space $\mathcal{D}$, which has dimension $|D|$. $\{\ket{z}:z \in D\}$ is the standard basis of $\mathcal{D}$. The domain $D$ is usually a feasible subset (following a specific set of constraints) of a larger configuration space. Now we define parameterized families of operators that act on $\mathcal{D}$.\par
\noindent We should be able to create a feasible initial state $\ket{s},s\in D$ efficiently from the $\ket{0}^{\otimes |D|}$ state. First, we apply to $\ket{s}$ the family of phase-separation operators $U_{P}(\gamma)$ that depend on the objective function $f$. Normally we define $U_{P}(\gamma)$ as $U_{P}(\gamma)=e^{-i\gamma H_{f}}$ where $H_{f}$ is the Hamiltonian corresponding to the objective function $f$ (we follow the techniques outlined in \cite{hadfield2018representation}). However, we could alter the definition to suit our needs. Next, we have the family of mixing-operators $U_{M}(\beta)$ which depend on $D$ and it's structure. $U_{M}(\beta)$ must preserve the feasible subspace, and provide transitions between all pairs of feasible spaces. A \qaoa{} circuit consists of $p$ alternating layers of $U_P(\gamma)$ and $U_M(\beta)$ applied to a suitable initial state $\ket{s}$.
\begin{equation}
    \ket{\gamma, \beta} = e^{-i\beta_{p}H_{m}}e^{-i\gamma_{p}H_{p}}\ldots e^{-i\beta_{1}H_{m}}e^{-i\gamma_{1}H_{p}}\ket{s}\label{gamma_beta_State}
\end{equation}
A computational basis measurement over the state $\ket{\gamma,\beta}$ returns a candidate solution state $\ket{z}$ having objective function value $f(z)$ with probability $\left|\bra{z}\ket{\gamma,\beta}\right|^{2}$. The goal of QAOA and \qaoa{} is to prepare a state $\ket{\gamma, \beta}$, from which we can sample a solution $z$ with a high value of $f(z)$.
\section{Designing the Quantum Alternating Operator Ansatz circuit for the Matching problem}
\noindent A matching $M$ in a graph $G(V,E)$ is a set of independent edges. Let us assume $G$ is undirected. We define a variable $x_e$ for every edge $e\in E$ and a constraint for every vertex $v\in V$. Consider the following integer linear program:
\begin{equation}
    \begin{split}
        &\text{maximize } \sum_{e\in{E}}x_{e}\\
        &\text{subject to}\\
        &\sum_{e \sim v}x_{e}\leq 1,\;\;\forall v\in V\\
        & x_{e} \in \{0,1\},\;\;\forall e\in E\\
    \end{split}\label{Max_Matching_ILP}
\end{equation}
Here $e\sim v$ means edge $e$ is incident on vertex $v$. The solution to the ILP in \eqref{Max_Matching_ILP} gives us the maximum matching for $G$.\par
\noindent Each individual qubit in the basis state corresponds to an edge in $G$. For example, in a rectangular graph or $C_4$ (cycle with 4 edges), the state $\ket{1010}$ denotes a matching containing the first and third edges only, and $\ket{0000}$ indicates the empty matching.\par
\noindent We consider two choices for our initial state $\ket{s}$. The first choice is the empty matching $\ket{0}^{\otimes |E|}$. The second choice is the $W_{1}$ state, which is a generalization of $W$ states as defined in \cite{PhysRevA.62.062314}. $\ket{W_1}$ is a uniform superposition over all states of Hamming Weight $1$. We note that both of these choices are feasible solutions to \eqref{Max_Matching_ILP}, and hence form valid matchings.\par
\noindent Our objective function $g(x) = \sum_{e\in{E}}\;x_{e}$ counts the number of edges in the matching. We map it to the Hamiltonian $H_g$ (using Hamiltonian composition rules from \cite{hadfield2018quantum}) described below.
\begin{equation}
    H_g = \sum_{e\in E}\frac{1}{2}(I-Z_{e})=\frac{|E|}{2}I-\frac{1}{2}\sum_{e\in E}Z_{e}\label{phase_hamil}
\end{equation}
In \eqref{phase_hamil}, $X$ and $Z$ denote the unitaries for Pauli-X and Pauli-Z respectively. $Z_e$ signifies that the Pauli-Z operator is applied to the $e$\textsuperscript{th} qubit. As discussed before, the family of phase-separation operators is diagonal in the computational basis. Our phase separation unitary is $U_{P}(\gamma) = e^{i\gamma H_g}$. We drop the constant term (since it affects the algorithm by a global phase) to get:
\begin{equation}
    U_{P}(\gamma) = e^{i\frac{\gamma}{2}\sum_{e\in E}Z_{e}}=\prod_{e\in E} e^{i\frac{\gamma}{2}Z_{e}} = \prod_{e\in E}R_{Z_{e}}(-\gamma)\label{phase-separation-unitary}
\end{equation}
\begin{definition}[Control Clause]
 The constraints are programmed into the control clause
 \begin{equation}
 f(e)=\prod_{\Tilde{e} \in \mathrm{nbhd}(e)} \overline{x_{\Tilde{e}}}    
 \end{equation}
 where $\mathrm{nbhd}(e)$ refers to the edges that are adjacent to $e$.\label{def:control_clauses}
\end{definition}
\noindent The mixing unitary is responsible for evolving our system from one feasible state to another feasible state. We achieve this by encoding the constraints from \eqref{Max_Matching_ILP} into the mixing Hamiltonian $M_{e} = f(e)X_{e}$, using control clauses. The corresponding unitary operator is:
\begin{equation}
    U_{M,e}(\beta) = e^{-i\beta M_{e}} = \bigwedge_{f(e)}\left(e^{-i\beta X_{e}}\right) = \bigwedge_{f(e)}R_{X_{e}}(\beta)\label{mixing_unitary}
\end{equation}
$R_{X_e}$ is the $X$-rotation gate applied to the qubit $e$. In \eqref{mixing_unitary}, $\bigwedge_{f(e)}R_{X_{e}}(\beta)$ signifies a multi-qubit controlled $X$-rotation gate, where the control on the $R_{X_e}$ unitary is the control clause $f(e)$ corresponding to qubit $e$. Equation \eqref{mixing_unitary} can be efficiently implemented using multi-qubit controlled rotation gates. 
\noindent In one round of the \qaoa{} algorithm, we apply the individual mixing unitaries to every qubit. The (consolidated) mixing unitary is mathematically represented as:
\begin{equation}
    U_{M}(\beta)= \prod_{e\in E}U_{M,e}(\beta)= \prod_{e\in E}e^{-i\beta M_{e}}=\prod_{e\in E}\bigwedge_{f(e)}R_{X_{e}}(\beta)\label{consolidated-mixing-unitary}
\end{equation}
Since the mixing unitaries as defined in \eqref{mixing_unitary} are not necessarily diagonal in the computational basis, the ordering of the unitaries in \eqref{consolidated-mixing-unitary} matters. We formally define the concept of fixed orderings and arbitrary orderings in Definition \ref{def:ordering}, when we prove results for $2$-regular graphs. At this point we also note that for $p=1$, the depth of our circuit is polynomial with respect to the input size (the number of edges $|E|$) as the Phase Separation unitary can be implemented in depth $1$, while the Mixing Unitary can implemented in depth $c\cdot|E|, c<5$.\par
\noindent We now prove that the mixing unitary preserves the feasibility of the input state.
\begin{figure*}[tbhp]
    \centering
    \begin{subfigure}{0.35\linewidth}
        \includegraphics[width=\linewidth]{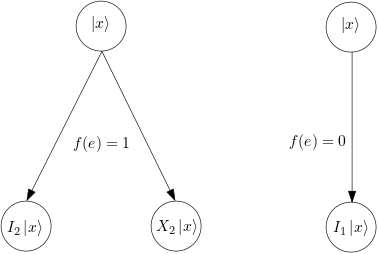}
        \caption{Construction tree branches}
        \label{fig:cases}
    \end{subfigure}
    \begin{subfigure}{0.6\linewidth}
        \includegraphics[width=\linewidth]{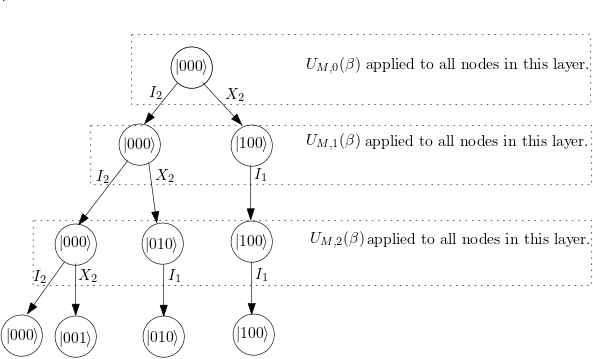}
        \caption{Construction tree for the $C_3$ graph (Triangle Graph)}
        \label{fig:construction_tree_c3}
    \end{subfigure}
    \hfill
    \caption{Construction Tree}
    \label{fig:construction_tree}
\end{figure*}
\begin{lemma}\label{lemma:feasibility_mixing}
The mixing operator $U_M(\beta)$ preserves the feasibility of the initial state $\ket{s}$. If $\ket{s}$ is feasible, then $U_M(\beta)\ket{s}$ is also feasible.
\end{lemma}
\begin{proof}
At any intermediate stage, let the quantum state be represented as $\ket{x}=\ket{x_1 x_2\ldots x_n}$. After applying an individual mixing unitary to its corresponding qubit $x_e$, the resulting state $U_{M,e}(\beta)\ket{x}$ can be expanded and written as
\begin{equation}
    \begin{split}
        U_{M,e}(\beta)\ket{x}&=\bigwedge_{f(e)}R_{X_{e}}\ket{x}\\&=f(e)R_{X_{e}}\ket{x}+\overline{f(e)}I\ket{x}\\
        &=\left(f(e)\cos(\beta/2)+\overline{f(e)}\right)\ket{x}\\ &-if(e)\sin(\beta/2)\ket{x_{1}\ldots \overline{x}_{e}\ldots x_{n}}
    \end{split}
     \label{mixing_feasibility}
\end{equation}
From \eqref{mixing_feasibility}, we see that we can have two cases:
\begin{enumerate}
    \item The control clause evaluates to 0, $f(e)=0$. This means that the output state is $\ket{x}$ itself.
    \item The control clause evaluates to 1, $f(e)=1$. This means that none of the edges adjacent to current edge $e$ is already selected as part of the matching. The resultant state is a superposition between a matching including current edge $e$, and a matching excluding current edge $e$. expand with respect to $x_e$. The resulting state is consistent with the constraints of \eqref{Max_Matching_ILP}.
\end{enumerate}
Hence we see that if $\ket{x}$ is a feasible state, then $U_{M,e}(\beta)\ket{x}$ is also feasible. We can now show by induction that the output state $U_M(\beta)\ket{s}$ will always be a superposition of feasible states, if and only if the initial state $\ket{s}$ is feasible.
\end{proof}
\noindent We note that in \eqref{mixing_feasibility}, each $U_{M,e}(\beta)$ unitary is composed of three separate unitaries. We rename these unitaries, as it makes most of the subsequent analysis easier.
\begin{definition}[Renaming the unitaries]\label{def:unitaries}
We rename the component unitaries of \eqref{mixing_feasibility} as follows:
\begin{enumerate}
    \item When $f(e)=0$, $U_{M,e}(\beta)\ket{x}=I\ket{x}$. We denote this as the $I_1$ unitary.
    \item When $f(e)=1$, $U_{M,e}(\beta)\ket{x}=\cos{(\beta/2)}\ket{x}-i\sin{(\beta/2)}X_e\ket{x}$.
    \begin{itemize}
        \item The $(\cos{\frac{\beta}{2}})I$ operator is denoted as the $I_2$ unitary, and 
    \item the $(\sin{\frac{\beta}{2}})X_e$ operator is denoted as the $X_2$ unitary. 
    \end{itemize}
\end{enumerate}
\end{definition}
\section{\qaoa{} with empty matching as the initial state}
\noindent The empty matching can be represented by the state $\ket{0}^{\otimes |E|}$. Using this as out initial state we derive a few results for our \qaoa{} setup. First we define the concept of a construction tree.
\begin{definition}[Construction Tree]
In the \qaoa{} setup for $p=1$, applying the $U_{M,e}(\beta)$ unitary on every qubit gives rise to at most two different branches of computation at every step. Thus the actions of the mixing unitary $U_{M}(\beta)$ can be represented as a binary tree having exactly $|E|+1$ layers. Each layer $i, 0<i<|E|$ corresponds to the possible actions we can take for the current edge $e_i$, conditioned on the actions we have taken on edges $e_0$ to $e_{i-1}$. We refer to the binary tree corresponding to a given \qaoa{} setup (for $p=1$) as its construction tree.
\end{definition}
We can see the concept of branches of computation in Figure \ref{fig:cases} and an example construction tree for the $C_3$ graph in Figure \ref{fig:construction_tree_c3}.
\begin{theorem}\label{thm:superposition_all_matchings}
Applying $|E|$ unitaries of type $U_{M,e}(\beta)\ket{x}$ in the $\qaoa{}_{p=1}$ setup with $\ket{0}^{\otimes |E|}$ as initial state, yields a superposition over all possible distinct matchings with non-zero amplitudes, in a graph with $|E|$ edges, where $\beta\in(0,\pi)$.
\end{theorem}
If the controls permit, the output of applying $U_{M,e}(\beta)$ on the current state $\ket{x}$ is a superposition of a matching including the edge $e$, and a matching excluding the edge $e$. There is always a branch of construction which allows us to pick the current edge, and the rest of the subtree is conditioned on this choice. Hence, if we start with the empty matching, for any matching, at least one of the branches of our construction tree always gives that particular matching. This shows us that if there exists a feasible matching, then there exists a branch of construction which gets to that matching in $|E|$ steps. Here, $|E|$ steps are necessary since the edges might be supplied to us in an arbitrary order. This is seen in Figure \ref{fig:construction_tree_c3}. We now prove this formally.
\begin{proof}
First, we show that all possible distinct matchings are present in the output state of our \qaoa{} setup for $p=1$. This proof is obtained by induction on the construction tree. We hypothesize that at the start of layer $i, i>0$ we have a superposition over all possible distinct matchings of size at most $i$ with non-zero amplitude, using edges $e_0$ to $e_{i-1}$. We can easily verify this for the base case $i=1$, where we have a superposition over two matchings with non-zero amplitude - a matching including edge $e_0$ and a matching excluding edge $e_0$. For the inductive step, let us assume that at the start of layer $k$ we have a superposition over all possible distinct matchings with non-zero amplitude of size at most $k$, using edges $e_0$ to $e_{k-1}$. Now we apply $U_{M,e_k}(\beta)$ to every node in this layer.
\begin{itemize}
    \item When $f(e)=0$, only the current matchings are carried forward to the next layer.
    \item When $f(e)=1$, we carry forward both the current matchings, and new matchings which are formed by union of the current matchings and edge $e_k$.
\end{itemize}
This exhaustively creates all distinct matchings of size at most $k+1$ with non-zero amplitude, using edges $e_0$ to $e_{k}$, since we have assumed that the induction step is true and all distinct matchings of size at most $k$ with non-zero amplitude were present at the beginning of the current layer. By principle of mathematical induction, at the end of $|E|$ layers, we have a superposition over all possible distinct matchings of size at most $|E|$ with non-zero amplitude, using all edges.
\end{proof}
\noindent Now, we define two concepts, and an important corollary of Theorem \ref{thm:superposition_all_matchings}.
\begin{definition}
The number of distinct $k$-matchings in a graph $G$ is given by the function $\Phi_k(G)$. We also define $\Phi(G)$ as
\begin{equation}\label{eq:number_of_matchings}
    \Phi(G) = \sum_{k=0}^{\nu(G)} \Phi_{k}(G)
\end{equation}
where $\nu(G)\leq\lfloor |E|/2\rfloor$ is the matching number of graph $G$.\label{def:phik_G}
\end{definition}
We can make the following observation from the construction tree
\begin{observation}\label{thm:number_of_matchings}
In the construction tree for Theorem \ref{thm:superposition_all_matchings} for $\textsc{\qaoa}_{p=1}$, we have exactly $\Phi(G)$ number of leaves, where each leaf corresponds to a distinct matching. We also have exactly $\Phi(G)$ number of branches, each ending in a distinct leaf.
\end{observation}
\begin{proof}
In ${p=1}$, every layer of the construction tree corresponds to all possible choices we make regarding one particular edge. Once an edge has been seen, we never go back to it again in the same iteration. Each branch of the computation represents a unique sequence of operators applied to the edges, since every subtree is conditioned on the choices taken on the earlier edges. Hence, there are no two branches in the construction tree producing the same output state. By Theorem \ref{thm:superposition_all_matchings}, we see that the output state is a superposition over all possible distinct matchings. Combining the two arguments gives us exactly $\Phi(G)$ number of branches, each ending in a distinct leaf.
\end{proof}
\section{\qaoa{} with $W_1$ state as the initial state}
\noindent $\ket{W_1}$ represents an uniform superposition over all matchings of size $1$ in the context of our \qaoa{} setup.
\begin{equation}
    \ket{W_1}=\frac{1}{\sqrt{|E|}}\left(\ket{10\ldots 0}+\ket{01\ldots 0}+\ldots+\ket{0\ldots01}\right)
\end{equation}
With the help of $\ket{W_1}$ states, we are able to eliminate the empty matching from the superposition of states in the output state. In this section we explore the behaviour of the output state, and show that we converge to a superposition over maximal matchings with non-zero amplitudes in expected number of iterations almost $|E|$.\par
\noindent At this point we must note that the definition of control clauses as given in Definition \ref{def:control_clauses} is not sufficient to demonstrate the superiority of using $\ket{W_1}$ states over $\ket{0}^{\otimes |E|}$ theoretically. Hence we put forward the following modification
\begin{definition}\label{def:modified_control}
We update the definition of $f_e(x)$ as given in Definition \ref{def:control_clauses} to include the current qubit in its own control set.
\begin{equation}
    f(e)=\overline{x_e}\cdot\prod_{\Tilde{e} \in \mathrm{nbhd}(e)} \overline{x_{\Tilde{e}}}
\end{equation}
Thus the control clause is set when neither the current edge, nor its adjacent edges are already part of a matching.
\end{definition}
\noindent The advantage offered by this small modification is significant (as seen in Theorem \ref{thm:w1_state}). We see that if we only evaluate the neighbourhood of $e$, then in the case where $\ket{x}$ represents a matching including the current edge (this is possible when the initial matching already contains the current edge, like in the $W_1$ state), the resulting state after applying $U_{M,e}$ only retains the current matching or reduces the size of the matching. If we use the updated control clause, then in both cases ($x$ contains/does not contain $e$), applying $U_{M,e}$ retains the current matching or increases the size of the matching.
\begin{theorem}\label{thm:w1_state}
Applying $|E|$ unitaries of type $U_{M,e}(\beta)$ in the $\textsc{\qaoa}_{p=1}$ setup with the $W_1$ state as initial state and the modified control set as given in Definition \ref{def:modified_control}, yields a superposition over all possible non-empty distinct matchings, in a graph with $|E|$ edges where $\beta\in(0,\pi)$. 
\end{theorem}
\begin{proof}
If we use the control sets from Definition \ref{def:control_clauses}, then there exists a possibility that the current edge might be flipped to $0$, giving rise to the possibility of the empty matching existing. With the control clauses of Definition \ref{def:modified_control}, we simply apply $I_1$ to the current edge if it was already in the initial matching. This means that we never get the empty matching in the output state. The rest of the proof follows the proof of Theorem \ref{thm:superposition_all_matchings}.
\end{proof}
\begin{figure}[htbp]
    \centering
    \includegraphics[width=\linewidth]{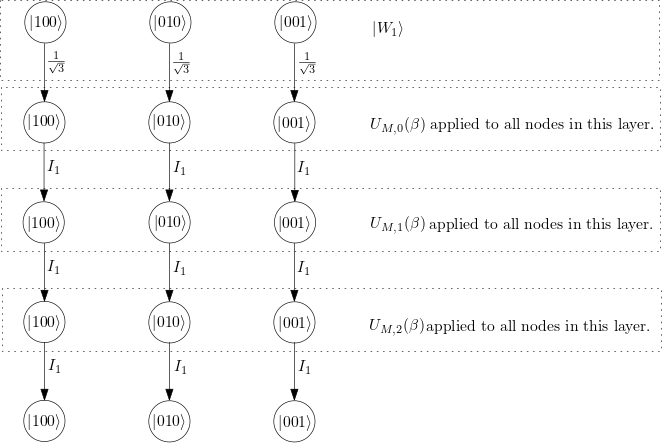}
    \caption{Construction Tree for $C_3$ using $\ket{W_1}$ as the initial state}
    \label{fig:c3_w1_construction}
\end{figure}
\begin{definition}
Let $\Phi^{+}(G)$ be defined as
\begin{equation}\label{eq:number_of_matchings+}
    \Phi^{+}(G) = \sum_{k=1}^{\nu(G)} k\cdot\Phi_{k}(G)
\end{equation}
where $\nu(G)\leq\lfloor |E|/2\rfloor$ is the matching number of graph $G$.
\end{definition}
\begin{observation}\label{thm:number_of_matchings_w1}
In the construction tree for Theorem \ref{thm:w1_state} for $\textsc{QAOA}_{p=1}$, we have exactly $\Phi^{+}(G)$ number of leaves, where each leaf corresponds to a distinct matching. We also have exactly $\Phi^{+}(G)$ number of branches, each ending in a distinct leaf.
\end{observation}
\begin{proof}
The construction tree for Theorem \ref{thm:w1_state} can be decomposed into $|E|$ different construction trees, corresponding to the $|E|$ different states in the initial $\ket{W_1}$ state. If we consider the construction tree corresponding to the $i$\textsuperscript{th} edge, then the output states of the $i$\textsuperscript{th} tree consists exactly of all matchings that include the $i$\textsuperscript{th} edge. Hence every distinct matching of size $k$ occurs is present exactly $k$ times in the output state when we consider all the $E$ construction trees together. Now following the arguments of Theorem \ref{thm:number_of_matchings}, we get that there are exactly $\Phi^+(G)$ number of states in the output of $\qaoa{}_{p=1}$, when we have $\ket{W_1}$ as the initial state.
\end{proof}

\begin{theorem}\label{thm:max_no_iterations}
In a graph with $|E|$ edges, with the modified control set from Definition \ref{def:modified_control}, $\beta\in[\pi/2,\pi)$ and $\ket{W_1}$ as the initial state, we can obtain an output state which is a superposition over all maximal matchings with non-zero amplitudes in $p\leq 2|E|$.
\end{theorem}
\begin{proof}
If $f(e)=1$, then the $X_2$ unitary is applied to the qubit $e$ with the probability ${\sin^{2}(\beta/2)}$. This means that on expectation, it will take $1/{\sin^{2}(\beta/2)}$ iterations for us to apply the $X_2$ unitary to the qubit $e$. Since Definition \ref{def:modified_control} ensures that the expected matching size of iteration $p$ is either greater than or equal to the expected matching size of iteration $p-1$, we see that we converge to a superposition over all maximal matchings in $p\leq\frac{1}{\sin^{2}(\beta/2)}|E|$. This concludes our proof since $\beta\in[\pi/2,\pi)$.
\end{proof}
\noindent The output state from Theorem \ref{thm:max_no_iterations} is a superposition over maximal matchings with non-zero amplitudes. Any further iterations of \qaoa{} depends on the phase separation operator only, and the mixing operator does not work. From here, Grover diffusion techniques may be used to increase the amplitude of the state with greater hamming weight, and using this in conjunction with good sampling techniques yields the maximum matching with high probability.
\section{\qaoa{} applied to $2$-regular graphs}
\noindent Until now we have been showing results for general graphs. However, in order to prove stronger bounds, we have to limit our focus to cycle graphs (\Cn) and \treg{} graphs. In this section we show that the expected matching size of the output state when using $\ket{W_1}$ as initial state is greater than the expected matching size of the output state when using $\ket{0}^{\otimes{|E|}}$ as initial state. Additionally we also show that the expected matching size of the output state when using $\ket{0}^{\otimes{|E|}}$ as initial state is greater than the expected matching size obtained from a uniform superposition over all matchings.\par
\noindent First, we would like to formally define the concept of fixed orderings and arbitrary orderings.
\begin{definition}[Fixed Ordering and Arbitrary Ordering]\label{def:ordering}
We define a cyclical ordering of a cycle graph $C_n$ as ordering the edges from $e_0$ to $e_{n-1}$ in a clockwise manner. A fixed ordering is a ordering of edges, in which the mixing unitary $U_M(\beta)$ acts on the edges in a cyclical order. If the edges are not supplied to the \qaoa{} algorithm in a cyclical order, we refer to this ordering of edges as an arbitrary ordering.
\end{definition}
\noindent We shall mention a few lemmas now, which can be shown via counting arguments:
\begin{lemma}\label{lemma:cycle_graph_k_mat}
The number of distinct $k$-matchings in a cycle graph \Cn{} is
\begin{equation}
    \Phi_k (\textsc{C}_n) = \frac{n}{n-k}\binom{n-k}{k}
\end{equation}
\end{lemma}
\begin{lemma}\label{lemma:path_graph_k_mat}
The number of distinct $k$-matchings in a path graph $\textsc{P}_n$ is
\begin{equation}
    \Phi_k(\textsc{P}_n) = \binom{n-k}{k}
\end{equation}
\end{lemma}
\begin{lemma}\label{lemma:component_graph_k_mat}
The number of distinct $k$-matchings in a Graph $G$ with $r$ components $(G_1,G_2,\ldots,G_r)$ is
\begin{equation}
    \Phi_k(G) = \prod_{i=1}^{r}\Phi_k(G_i)
\end{equation}
\end{lemma}
\begin{theorem}\label{thm:exp_mat_w1_empty}
In a $2$-regular graph with $|E|>16$ edges, with fixed ordering, and $\beta\in\left({\pi/2},{\pi}\right]$, the expected matching size of the output of $\textsc{\qaoa}$ for ${p=1}$ with $\ket{W_1}$ as the initial state is greater than the expected matching size size of the output of $\textsc{\qaoa}$ for ${p=1}$ with $\ket{0}^{\otimes |E|}$ state as the initial state. 
\end{theorem}
\begin{proof}
The construction tree of $\textsc{\qaoa}$ for ${p=1}$ with $\ket{W_1}$ as the initial state, can be represented as $|E|$ parallel construction trees of depth $E+1$ each having a distinct matching of size $1$ in the first layer. Let us consider the tree $T_i$, where the initial matching is $\{e_i\}$. Using the arguments of Theorems \ref{thm:superposition_all_matchings} and \ref{thm:w1_state} we see that $T_i$ produces exactly all matchings which contain the edge $e_i$. Similarly from Observation \ref{thm:number_of_matchings_w1}, we can argue that every matching of size $k$, is produced in exactly $k$ construction trees.\par
\noindent Let the amplitude of an arbitrary matching $M$ of size $k$ in the output of $\textsc{\qaoa}$ for ${p=1}$ with $\ket{0}^{\otimes |E|}$ state as the initial state be denoted as $\alpha_{M,0}$. Let the amplitude of $M$ for the $\ket{W_1}$ case be denoted as $\alpha_{M,W}$.\par
\noindent In the $\ket{W_1}$ case, there are $k$ trees which give rise to the matching $M$. In one of these trees $T_i$, the edge $e_i$ is already included in the initial matching. Let the amplitude of $M$ for this tree be denoted as $\alpha_{M,W,d}$. Since we use the modified control clauses, it means that we don't have to apply
\begin{itemize}
    \item The $X_2$ unitary on edge $e_i$ to include it in the matching $M$.
    \item The $I_2$ unitary on all edges in $\mathrm{nbhd}(e_i)$ that appears before $e_i$ in the fixed ordering, to exclude them from the matching.
\end{itemize}
In both of these cases we simply use $I_1$ unitaries which does not change the amplitude of the state. Let $d$ be the number of edges in $\mathrm{nbhd}(e_i)$, that appear before $e_i$ in the fixed ordering. We know that $d\in[0,k-1]$, as a particular matching of size $k$ is generated in $k$ out of the $|E|$ construction trees. Hence $\alpha_{M,W,d}$ can be expressed in terms of $\alpha_{M,0}$ as
\begin{equation}
        \alpha_{M,W,d}=\frac{\alpha_{M,0}}{\sqrt{|E|}\sin{(\beta/2)}\cos{}^{d}(\beta/2)}
\end{equation}
The $1/\sqrt{E}$ term comes from the nature of the $\ket{W_1}$ state. The probability of obtaining $M$ in the $W_1$ case is given as
\begin{equation}
    \alpha_{M,W}^{2}=\sum_{d=0}^{k-1}\alpha_{M,W,d}^{2}
\end{equation}
We can express this in terms of the probability of obtaining $M$ in the $\ket{0}^{\otimes |E|}$ case as
\begin{equation}
        \label{prob_comp_sum_w_0}
        \alpha_{M,W}^{2}=\sum_{d=0}^{k-1}\frac{\alpha_{M,0}^{2}}{|E|\sin^{2}{(\beta/2)}\cos^{2d}(\beta/2)}  
\end{equation}
For the range $\beta\in\left({\pi/2},\pi\right]$ we have
\begin{equation}
\label{prob_comp_w_0}
    \alpha_{M,W}^{2}\geq\frac{2^{k+1}-2}{|E|}\cdot{\alpha_{M,0}^{2}}
\end{equation}
In \eqref{prob_comp_w_0}, $2^{k+1}-2>|E|$ occurs, when $k\geq\log_{2}\left({|E|}+2\right)$. Therefore in the interval $\left[\log_{2}\left({|E|}+2\right),\nu(G)\right]$, we have $\alpha_{M,W}^{2}>\alpha_{M,0}^{2}$.\par
\noindent Let us define a random variable $X$, which denotes the size of the matching in a cycle graph. We calculate the expected matching size as
\begin{equation}
\begin{split}
    \ex{X}=&\underset
        {\text{Case A}}
        {   \sum_{k_A=0}^{\log_{2}\left({|E|}+2\right)-1}k_A\mathbb{P}[X=k_A]
        }\\&+\\&\underset{\text{Case B}}
    {
        \sum_{k_B=\log_{2}\left({|E|}+2\right)}^{\nu(G)}k_B\mathbb{P}[X=k_B]
    }
\end{split}
\end{equation}
Now we compare the expected matching size in the two cases. For Case A we upperbound $\ex{X}_{\ket{0}^{\otimes |E|}}-\ex{X}_{\ket{W_1}}-$ as:
\begin{equation}\label{uppercaseA}
    \begin{split}
    &{\sum_{k_A=0}^{\log_{2}\left({|E|}+2\right)-1}k_A\cdot\left(\mathbb{P}[X=k_A]_{\ket{0}^{\otimes |E|}}-\mathbb{P}[X=k_A]_{\ket{W_1}}\right)}\\
    &={\sum_{k_A=0}^{\log_{2}\left({|E|}+2\right)-1}k_A\cdot\sum_{M,|M|=k_A}\left(\alpha_{M,0}^{2}-\alpha_{M,W}^{2}\right)}\\
    &\text{To upperbound }\left(\alpha_{M,0}^{2}-\alpha_{M,W}^{2}\right)\text{ put }k_A=0\text{ in }\alpha_{M,W}^{2}
    \\
    &\leq \sum_{k_A=0}^{\log_{2}\left({|E|}+2\right)-1}k_A\cdot\sum_{M,|M|=k_A}\alpha_{M,0}^{2}
    \end{split}
\end{equation}
For Case B we lowerbound $\ex{X}_{\ket{W_1}}-\ex{X}_{\ket{0}^{\otimes |E|}}$ as:
\begin{equation}\label{lowercaseB}
    \begin{split}
    &{\sum_{k_B=\log_{2}\left({|E|}+2\right)}^{\nu(G)}k_B\cdot\left(\mathbb{P}[X=k_B]_{\ket{W_1}}-\mathbb{P}[X=k_B]_{\ket{0}^{\otimes |E|}}\right)}\\
    &={\sum_{k_B=\log_{2}\left({|E|}+2\right)}^{\nu(G)}k_B\cdot\sum_{M,|M|=k_B}\left(\alpha_{M,W}^{2}-\alpha_{M,0}^{2}\right)}\\
    &\text{To lowerbound }\left(\alpha_{M,W}^{2}-\alpha_{M,0}^{2}\right)\text{, put }\\
    &k_B=\log_{2}\left({|E|}+2\right)\text{ in }\alpha_{M,W}^{2}
    \\
    &\geq \sum_{k_B=\log_{2}\left({|E|}+2\right)}^{\nu(G)}k_B\cdot\sum_{M,|M|=k_B}\left(\frac{4|E|+6}{|E|}\cdot\alpha_{M,0}^{2}-\alpha_{M,0}^{2}\right)\\
    &> \sum_{k_B=\log_{2}\left({|E|}+2\right)}^{\nu(G)}k_B\cdot\sum_{M,|M|=k_B}\left(4\cdot\alpha_{M,0}^{2}-\alpha_{M,0}^{2}\right)\\
    &= \sum_{k_B=\log_{2}\left({|E|}+2\right)}^{\nu(G)}k_B\cdot3\cdot\sum_{M,|M|=k_B}\alpha_{M,0}^{2}
    \end{split}
\end{equation}
Since the sum of $\alpha_{M,0}^{2}$ over all possible matchings is $1$, we know that
\begin{equation}\label{point1}
    \sum_{M,|M|=k}\alpha_{M,0}^{2}\leq 1
\end{equation}
We set this sum to $1$. We always have
\begin{equation}\label{point2}
    k_B>k_A
\end{equation}
When $|E|>16$, we have
\begin{equation}\label{point3}
    \underset{\text{Number of terms in Case B}}{\nu(G)-\log_{2}(|E|+2)} > \underset{\text{Number of terms in Case A}}{\log_{2}(|E|+2)}
\end{equation}
From \eqref{point1}, \eqref{point2}, and\eqref{point3} we see that the lower bound obtained in \eqref{lowercaseB} is greater than the upper bound obtained in \eqref{uppercaseA}. Hence, we have 
\begin{equation}
        \ex{X}_{\ket{W_1}}>\ex{X}_{\ket{0}^{\otimes |E|}},\;\;\; |E|>16
\end{equation}
\noindent We know that $2$-regular graphs are composed of disconnected components, where each component is a cycle graph. Let us define a random variable $Y$, which denotes the size of the matching in a $2$-regular graph $G$. We also define random variables $Y_i$ for all the $r$ disconnected components of $G$, which denotes the size of the matching in $G_i$. Using linearity of expectation over $\ex{Y_i}$, we have
\begin{equation}
    \begin{split}
        \ex{Y_i}_{\ket{W_1}}&>\ex{Y_i}_{\ket{0}^{\otimes |E|}}\;\;\forall i\in\{1,2,\ldots,r\}\\
        \implies \ex{Y}_{\ket{W_1}}&>\ex{Y}_{\ket{0}^{\otimes |E|}},\;\;\;|E|>16
    \end{split}
\end{equation}
This concludes our proof.
\end{proof}
\begin{figure*}[bpht]
    \centering
    \includegraphics[width=\linewidth]{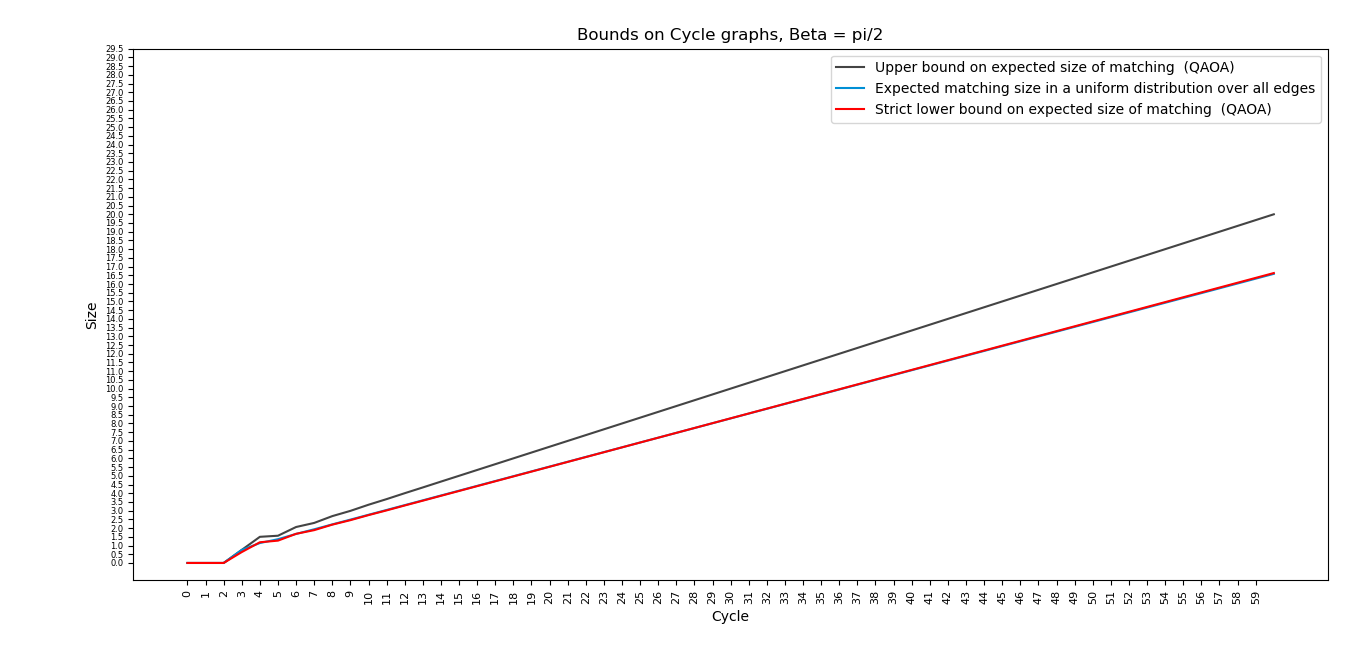}
    \caption{This figure corresponds to Theorem \ref{thm:qaoa_greater_uniform}. The upper and lower bounds for $\ex{X}_{\qaoa{}}$ vs $\ex{X}_{\texttt{uniform}}$ for $\beta=\pi/2$ on Cycle Graphs having up to 59 edges.}
         \label{fig:theorem_bounds}
\end{figure*}

\begin{theorem}\label{thm:qaoa_greater_uniform}
Let us consider a \qaoa{} for a cycle graph $C_n, n>6$, $p=1$, fixed ordering, empty initial matching, and $\beta\in\left[\frac{\pi}{2},\pi\right)$. The expected matching size of the \qaoa{} output state is greater than the expected matching size obtained from a uniform distribution over all matchings.
\end{theorem}
\begin{figure}[htbp]
    \centering
    \includegraphics[width=\linewidth]{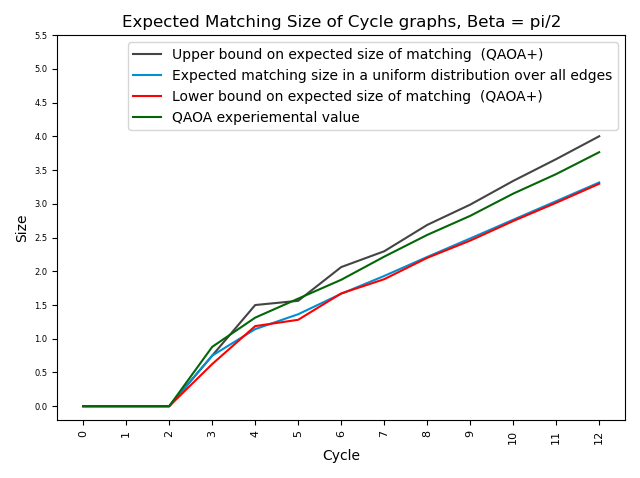}
         \caption{This figure corresponds to Theorem \ref{thm:qaoa_greater_uniform}. Expected matching size of \qaoa{} vs uniform distribution for $\beta=\pi/2$ on Cycle Graphs up to 12 vertices, using \texttt{qasm\_simulator}. Figure \ref{fig:experiment_bounds} is a zoomed-in version (along with experimental value obtained on \texttt{qasm\_simulator}) of Figure \ref{fig:theorem_bounds}.}
         \label{fig:experiment_bounds}
\end{figure}
\begin{figure*}[htbp]
     \centering
     \includegraphics[width=\textwidth]{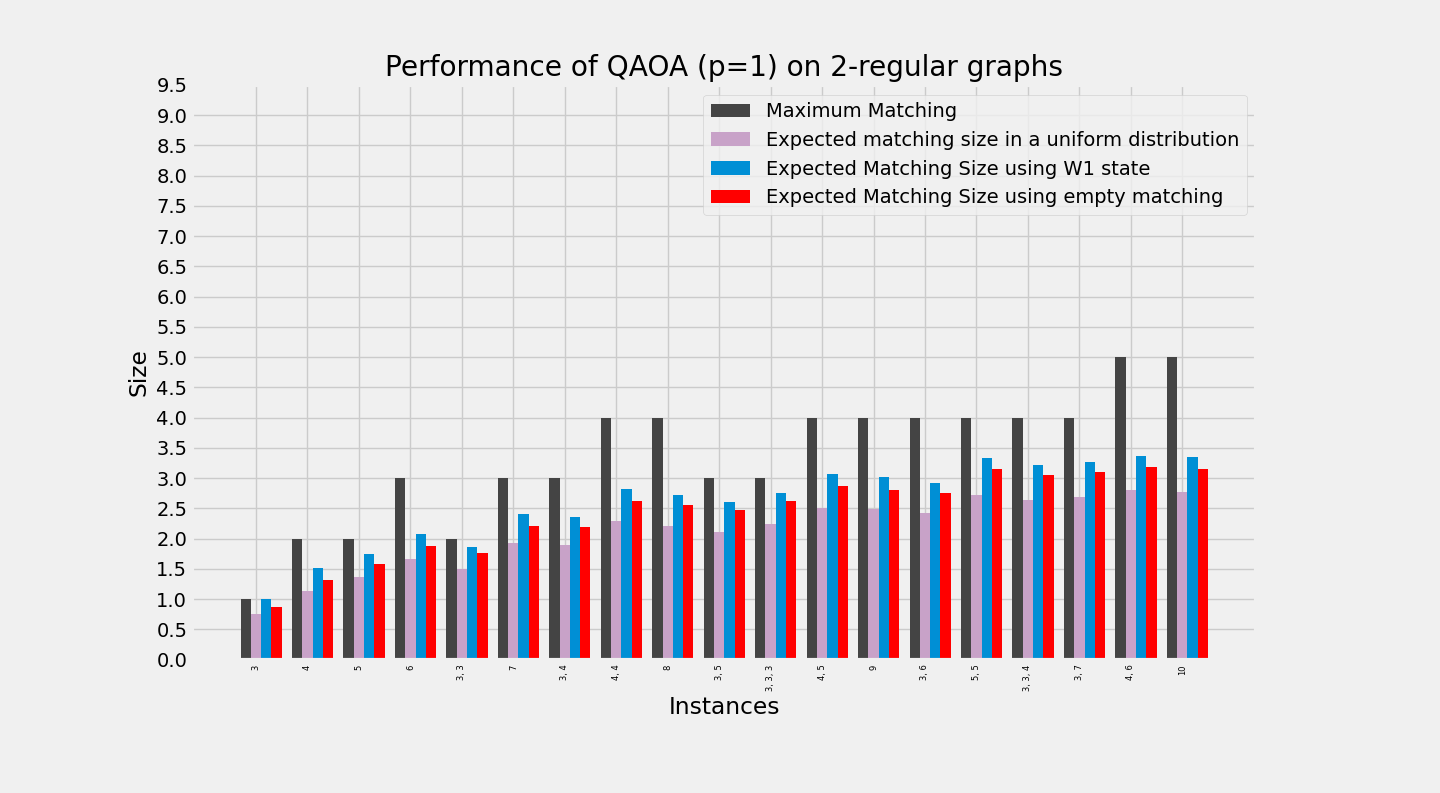}
     \caption{Experimentally demonstrating Theorem \ref{thm:exp_mat_w1_empty}, Theorem \ref{thm:qaoa_greater_uniform}, Corollary \ref{corr:empty_state_two_reg}, and Corollary \ref{corr:w1_state_two_reg} for $2$-regular graphs up to $10$ vertices, on the \texttt{qasm\_simulator} backend of QISKIT.}
     \label{fig:w1_empty_two_reg}
\end{figure*}
\noindent\textbf{Note:} We were unable to find a completely theoretical proof for Theorem \ref{thm:qaoa_greater_uniform}. The strict lower bound for $\ex{X}_{\qaoa{}}$ and the value for $\ex{X}_{\texttt{uniform}}$ have asymptotically identical curves. We obtained the proof for the theorem by plotting the values in equations \eqref{prob_quantum} and \eqref{expect_uniform_classical}. The results are shown in Figure \ref{fig:theorem_bounds}, for Cycle Graphs up to 50 vertices.
\begin{proof}
We want to calculate the expected size of matching in the output of \qaoa{}($p=1$), with $\ket{0}^{\otimes{|E|}}$ as the initial state. Let us define a random variable $X$, which denotes the size of the matching. We know that $\ex{X}=\sum_{k=0}^{\nu(G)}k\;\mathbb{P}[X=k]$. The value of $\mathbb{P}[X=k]$ can be obtained by squaring the amplitudes of the leaves in the construction tree of the \qaoa{} circuit as given in Theorem \ref{thm:number_of_matchings}, and adding together the probabilities of all $\Phi_k(G)$ leaves. Let $M_k$ denote a matching of size $k$, and $e_{\ell}$ denote the last edge. In order to calculate the probabilities of $k$-matchings in a cycle graph under the proposed \qaoa{} setup, we have to consider two cases:
\begin{enumerate}
    \item Case \textbf{a}: $\mathbf{e_{\ell}\notin M_k}$. This transforms our underlying graph into a path graph of $n$ vertices and edges. If the first edge is included in the matching, then for the last edge we have $f(e)=0$, leading to an $I_1$ unitary. If the first edge is not included the underlying graph is transformed into a path graph of $n-1$ vertices. The rest of the $n-2k$ unitaries will be $I_2$ unitaries, which contributes a total probability of $\cos^{2n-4k}{\beta/2}$. If the first edge is included, there are at most $n-2k$ $I_2$ unitaries. This analysis is recursive, so in order to make the calculation easier we lower bound the expectation. Let $s=\sin^2{\beta/2}$ and $c=\cos^2{\beta/2}$. Using Lemma \ref{lemma:path_graph_k_mat}, we have
    \begin{equation}
        \label{last_edge_not_in_matching}
        \begin{split}
            a &> s^{k}c^{n-2k}\left(\binom{n-k-1}{k}+\binom{n-k-1}{k-1}\right)\\
            \implies a&>\binom{n-k}{k}s^{k}c^{n-2k}
            \end{split}
    \end{equation}
    \item Case \textbf{b}: $\mathbf{e_{\ell}\in M_k}$. Then we have to count the number of distinct $(k-1)$-matchings path graph of $n-2$ vertices, which is $\phi_{k-1}(\textsc{P}_{n-2}) = \binom{n-k-1}{k-1}$. There are $k-1$ pairs of $X_2$ and $I_1$ unitaries, and one extra $X_2$ unitary for the last edge. The total contribution to the probability is $\sin^{2k}{\beta/2}$. The rest of the $n-(2k-1)$ edges are hit with the $I_2$ unitary, which contributes a total probability of $\cos^{2n-4k+2}{\beta/2}$. We again assign $s=\sin^2{\beta/2}$ and $c=\cos^2{\beta/2}$. Using Lemma \ref{lemma:path_graph_k_mat}, we have
    \begin{equation}\label{last_edge_in_matching}
            b = \binom{n-k-1}{k-1}s^{k}c^{n-2k+1}
            <\frac{k}{n-k}\binom{n-k}{k}s^{k}c^{n-2k}
    \end{equation}
\end{enumerate}
Combining \eqref{last_edge_not_in_matching} and \eqref{last_edge_in_matching}, for $n>5$:
\begin{equation}
        s^{k}\cdot c^{n-2k}\cdot \Phi_k(G) \left(1-\frac{s\cdot k}{n}\right)<\mathbb{P}[X=k]<s^{k}\cdot c^{n-2k}\cdot\Phi_k(G)\label{bound_prob}
\end{equation}
which gives us the lower bound for the expectated matching size as
\begin{equation}\label{bound_expectation}
    \ex{X}_{\qaoa{}}>\sum_{k=0}^{\nu(G)}s^{k}\cdot c^{n-2k}\cdot k\cdot\Phi_k(G)\left(1-\frac{s\cdot k}{n}\right)
\end{equation}
When $\beta=\pi/2$, we have the lower bound of \eqref{bound_prob} transform into:
\begin{equation}
        \ex{X}_{\qaoa{}}>\sum_{k=0}^{\nu(G)}k\cdot\Phi_k(G)\cdot\left(\frac{1}{2}\right)^{n-k}\cdot\left(1-\frac{k}{2n}\right)\label{prob_quantum}
\end{equation}
Let $M(G)$ be the set of matchings of graph $G$. Let us consider the uniform distribution over $M(G)$. When $G$ is $C_n$, we can calculate the probability of obtaining a matching of size $k$ as
\begin{equation}
    \mathbb{P}[X=k]_{\texttt{uniform}}=\frac{\Phi_k(G)}{\Phi(G)}
    \label{prob_uniform_classical}
\end{equation}
and the expected matching size as
\begin{equation}
    \ex{X}_{\texttt{uniform}}=\sum_{k=0}^{\nu(G)}k\cdot \mathbb{P}[X=k]
    =\sum_{k=0}^{\nu(G)}\frac{k\cdot\Phi_k(G)}{\Phi(G)}\label{expect_uniform_classical}
\end{equation}
As noted earlier, we get the theorem from plotting and comparing the values of \eqref{prob_quantum} and \eqref{expect_uniform_classical}.
\end{proof}
\noindent It can be noted that we may improve the bounds in \eqref{bound_expectation} at the expense of making the analysis more complicated, by continuing to analyze the recursive structure of the construction trees. Also we can further increase the values obtained in \eqref{prob_quantum}, by using a value of $\frac{\pi}{2}<\beta<\pi$.
\begin{corollary}[Theorem \ref{thm:qaoa_greater_uniform}]\label{corr:empty_state_two_reg}
Let us consider a \qaoa{} for a $2$-regular Graph, $p=1$, fixed ordering, empty initial matching, and $\beta\in\left[\frac{\pi}{2},\pi\right)$. The expected matching size of the output state is greater than the expected matching size obtained from a uniform distribution over all matchings.
\end{corollary}
\begin{proof}
The proof follows directly from using linearity of expectation on Theorem \ref{thm:qaoa_greater_uniform}. In a \treg{} graph, every component is a Cycle Graph. The expected matching size of the entire graph is a sum over expected matching size of the individual components. Our theorem follows.
\end{proof}
\begin{corollary}[Theorem \ref{thm:qaoa_greater_uniform}]\label{corr:w1_state_two_reg}
Let us consider a \qaoa{} for a $2$-regular Graph, $p=1$, fixed ordering, $W_1$ state as the initial matching, and $\beta\in\left[\frac{\pi}{2},\pi\right)$. The expected matching size of the output state is greater than the expected matching size obtained from a uniform distribution over all matchings.
\end{corollary}
\begin{proof}
The proof follows directly from Theorem \ref{thm:qaoa_greater_uniform} and Theorem \ref{thm:exp_mat_w1_empty}.
\end{proof}
\section{Conclusion and Future Work}
In this work we have seen how the Quantum Alternating Operator Ansatz (and by extension the Quantum Approximate Optimization Algorithm) framework can be applied to the graph matching problem. We have argued the merits of using $W_1$ states over the trivial empty state as the initial state and showed that a choice of initial state matters in terms of the output. We have also obtained a small amount of experimental validation for our various theoretical claims.\par
\noindent A possible future direction of this work would be investigating sampling techniques which produce the Maximum Matching from a superposition over all maximal matchings with non-zero amplitudes. Designing efficient samplers on the output state itself might have both algorithmic and complexity theoretic importance since counting problems with respect to matchings is \textsc{\#P}-hard as proved in \cite{valiant1979complexity}. Other future directions include theoretically improving the lower bounds of \eqref{prob_quantum}, and proving similarly strong results for more general classes of graphs.
\input{main.bbl}

\end{document}

%% file: main.bbl

%% file: main.bbl
\begin{thebibliography}{10}
\providecommand{\url}[1]{#1}
\csname url@samestyle\endcsname
\providecommand{\newblock}{\relax}
\providecommand{\bibinfo}[2]{#2}
\providecommand{\BIBentrySTDinterwordspacing}{\spaceskip=0pt\relax}
\providecommand{\BIBentryALTinterwordstretchfactor}{4}
\providecommand{\BIBentryALTinterwordspacing}{\spaceskip=\fontdimen2\font plus
\BIBentryALTinterwordstretchfactor\fontdimen3\font minus
  \fontdimen4\font\relax}
\providecommand{\BIBforeignlanguage}[2]{{%
\expandafter\ifx\csname l@#1\endcsname\relax
\typeout{** WARNING: IEEEtran.bst: No hyphenation pattern has been}%
\typeout{** loaded for the language `#1'. Using the pattern for}%
\typeout{** the default language instead.}%
\else
\language=\csname l@#1\endcsname
\fi
#2}}
\providecommand{\BIBdecl}{\relax}
\BIBdecl

\bibitem{farhi2014quantum}
E.~Farhi, J.~Goldstone, and S.~Gutmann, ``A quantum approximate optimization
  algorithm,'' 2014.

\bibitem{farhi2014lin}
E.~{Farhi}, J.~{Goldstone}, and S.~Gutmann, ``A quantum approximate
  optimization algorithm applied to a bounded occurrence constraint problem,''
  2014.

\bibitem{farhi2016quantum}
E.~Farhi and A.~W. Harrow, ``Quantum supremacy through the quantum approximate
  optimization algorithm,'' 2016.

\bibitem{Harrow_2017}
\BIBentryALTinterwordspacing
A.~W. Harrow and A.~Montanaro, ``Quantum computational supremacy,''
  \emph{Nature}, vol. 549, no. 7671, p. 203–209, Sep 2017. [Online].
  Available: \url{http://dx.doi.org/10.1038/nature23458}
\BIBentrySTDinterwordspacing

\bibitem{Hadfield_2019}
\BIBentryALTinterwordspacing
S.~Hadfield, Z.~Wang, B.~O’Gorman, E.~Rieffel, D.~Venturelli, and R.~Biswas,
  ``From the quantum approximate optimization algorithm to a quantum
  alternating operator ansatz,'' \emph{Algorithms}, vol.~12, no.~2, p.~34, Feb
  2019. [Online]. Available: \url{http://dx.doi.org/10.3390/a12020034}
\BIBentrySTDinterwordspacing

\bibitem{edmonds1965paths}
J.~Edmonds, ``Paths, trees, and flowers,'' \emph{Canadian Journal of
  mathematics}, vol.~17, pp. 449--467, 1965.

\bibitem{ambainis2005quantum}
A.~Ambainis and R.~Spalek, ``Quantum algorithms for matching and network
  flows,'' 2005.

\bibitem{valiant1979complexity}
L.~G. Valiant, ``The complexity of enumeration and reliability problems,''
  \emph{SIAM Journal on Computing}, vol.~8, no.~3, pp. 410--421, 1979.

\bibitem{hadfield2018representation}
S.~Hadfield, ``On the representation of boolean and real functions as
  hamiltonians for quantum computing,'' 2018.

\bibitem{PhysRevA.62.062314}
\BIBentryALTinterwordspacing
W.~D\"ur, G.~Vidal, and J.~I. Cirac, ``Three qubits can be entangled in two
  inequivalent ways,'' \emph{Phys. Rev. A}, vol.~62, p. 062314, Nov 2000.
  [Online]. Available:
  \url{https://link.aps.org/doi/10.1103/PhysRevA.62.062314}
\BIBentrySTDinterwordspacing

\bibitem{hadfield2018quantum}
S.~Hadfield, ``Quantum algorithms for scientific computing and approximate
  optimization,'' 2018.

\end{thebibliography}
